\documentclass[aps,prl,reprint,floatfix,longbibliography]{revtex4-2} 

\usepackage{amsmath}
\usepackage{amssymb}
\usepackage{bm}
\usepackage{graphicx}
\usepackage{placeins}
\usepackage{braket}
\usepackage{bbold}
\usepackage[colorlinks, linkcolor=blue, citecolor=blue, urlcolor=blue, breaklinks]{hyperref}
\usepackage{xcolor}
\usepackage{upgreek}  

\usepackage{mathtext}

\newcommand {\degree} {^{\circ}}

\newcommand  {\laser} {\mathrm{laser}}
\newcommand {\at} {\mathrm{at}}

\newcommand {\lat} {\mathrm{lat}}
\newcommand {\kB} {k_\mathrm{B}}
\newcommand {\GR} {\Gamma_\mathrm{R}}
\newcommand {\Rss} {R_\mathrm{ss}}
\newcommand {\inphyni} {Universit\'e C\^ote d'Azur, CNRS, Institut de Physique de Nice, France}

\begin{document}

\title{Temporal dynamics in the Bragg reflection of light by cold atoms:\\flash effect and superradiant decay}
\author{S. Asselie}
\author{J.-M. Nazon}
\author{R. Caldani}
\author{C. Roux-Spitz}
\author{W. Guerin}
\email{william.guerin@univ-cotedazur.fr}
\affiliation{\inphyni}

\begin{abstract}
We study the temporal dynamics of light interacting with a one-dimensional lattice of cold atoms. In such a system, a photonic band gap opens up, yielding an efficient Bragg reflection for an incident field incoming with the right angle and detuning. Here, we report two new effects appearing in the Bragg reflection. First, for some detunings, there is a ``flash'', i.e., a transient increase of the reflected intensity when the incident field is switched off. Second, the subsequent extinction of the reflected field is clearly superradiant, with decay rates up to height times the natural decay rate of the atomic excited state. Numerical simulations are in qualitative agreement with the observations, which can be explained by a classical photonic model. Our results are a step towards exploiting this photonic band gap in atomic systems for quantum-optical applications.
\end{abstract}

\date{\today}

\maketitle


Collective effects in light-atom interaction are of current high interest in the cold-atom community \cite{Guerin:2017a, Chang:2018, Reitz:2022, Sheremet:2023, Jen:2025}, for potential applications to quantum technologies (clocks \cite{Chang:2004}, quantum memories \cite{Asenjo:2017}, light harvesting \cite{Celardo:2014b, MorenoCardoner:2019}, novel lasing sources \cite{Guerin:2008, Bohnet:2012, Schilke:2012a, Baudouin:2013b, Holzinger:2020}, etc.) as well as for fundamental questions, for instance understanding the refractive index of gases \cite{Andreoli:2021}, or their spectral \cite{Javanainen:2016, Jennewein:2018, Vatre:2024} and temporal responses \cite{Guerin:2023}.

A lot of experimental work has been done using fully disordered samples, either in the macroscopic dilute regime (when the sample size and average distance between atoms are much larger than the wavelength $\lambda$), in which superradiance and subradiance have been observed \cite{Goban:2015, Guerin:2016a, Araujo:2016, Roof:2016}, or using small and dense clouds (size on the order of $\lambda$), in which superradiance and subradiance have been observed as well, with also spectral shifts \cite{Corman:2017, Jennewein:2018, Ferioli:2021a, Glicenstein:2022}. Another approach is to engineer the optical response by tailoring the atomic arrangement using microscopic, ordered ensembles containing a limited number of atoms. In spite of a huge theoretical production (see, e.g., ref. \cite{Jen:2025}), very few experiments were performed in this regime, most notably one-dimensional (1D) chains near nanofibers \cite{Corzo:2016, Solano:2017, Pennetta:2022} or in free space \cite{Glicenstein:2020,Hofer:2024}, and a mirror made of a single 2D atomic layer \cite{Rui:2020, Srakaew:2023}.

In this Letter, we present an alternative experimental platform, which is a hybrid system between order and disorder: we consider a macroscopic and dilute cold-atomic sample trapped in a one-dimensional conservative lattice. There is thus a long-range order (periodicity) in one direction but each atomic ``pancake'' in the array is dilute and disordered. As initially proposed by Deutsch \emph{et al.} with near-resonant lattices \cite{Deutsch:1995}, such a system creates a photonic band gap along the direction of the lattice, i.e. a range of optical frequency that cannot propagate in the medium \cite{Joannopoulos}. Incoming light is thus reflected off the medium. Such Bragg reflection has been experimentally realized much later using a far-detuned lattice, for which it was shown that a reflectivity up to 80\% could be obtained by a careful tuning of the lattice periodicity \cite{Schilke:2011}. In this previous work, however, only the steady-state photonic properties have been investigated. 

Here, we study the \emph{dynamical} properties of the Bragg reflection. In particular, we report two interesting effects observed at the switch-off of the incident probe beam. First, there is a nonintuitive \emph{optical flash}, similar to the one occurring in the coherent transmission through disordered samples \cite{Chalony:2011, Kwong:2014, Kwong:2015}. Second, the decay of the Bragg reflection is \emph{superradiant}, with rates a few times faster than the natural decay rate $\Gamma_0$ of the atomic excited state. 
The experimental observations are in good qualitative agreement with a simple 1D transfer-matrix model for the light propagation through the atomic layers.



\begin{figure}[t]
\centering\includegraphics{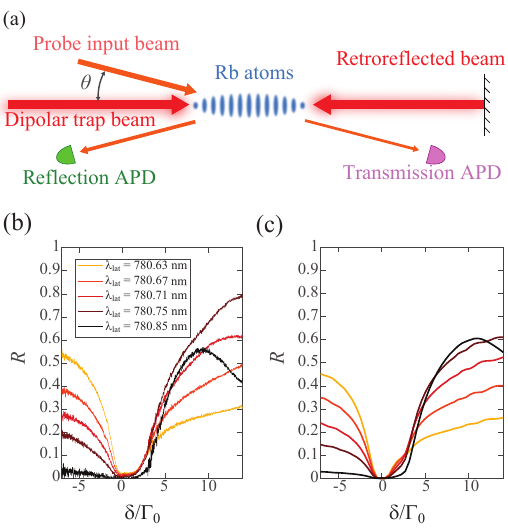}
\caption{(a) Simplified scheme of the setup. Cold atoms from a MOT are trapped in a retroreflected dipole trap forming a 1D lattice, and probed with a weak beam at a small angle $\theta$ from the lattice axis. The transmission $T$ and reflection $R$ are measured with avalanche photodiodes (APDs). (b) Bragg reflection spectra taken for different lattice wavelength $\lambda_\lat$. (c) Corresponding simulations. There is no free parameter except for the atom number (see main text). We also take into account the absorption near resonance due to the residual atoms from the MOT, which is not negligible for the short holding time in the lattice (here 30 ms) used to maximize the reflection.}
\label{fig.setup}
\end{figure}

Before discussing the results in detail, let us first describe the experimental setup and the sample parameters, which closely resemble the ones of Ref. \cite{Schilke:2011}. We load a 1D optical lattice from a compressed magneto-optical trap (MOT) of $^{87}$Rb. The lattice is made from a retro-reflected continuous tunable (titanium-sapphire) laser focused at the MOT position [Fig.\,\ref{fig.setup}(a)]. We use about 1\,W of optical power and the beam waist is $w_\lat \approx 380\,\upmu$m. The wavelength of the lattice $\lambda_\lat$ can be changed in order to finely tune the Bragg condition, which is related to the angle of the probe beam from the lattice axis, chosen at $\theta \approx 1.9\degree$. Using the $D_2$ line of Rb at $\lambda=780.24$\,nm, this corresponds to the geometric Bragg condition $\lambda_\lat = \lambda/\cos(\theta) \approx 780.67$\,nm. However, the true Bragg condition involves the wavelength \emph{in the medium}, i.e., taking into account the mean refractive index, such that it is more favorable to tune the lattice slightly farther from resonance, which yields a better Bragg reflection with asymmetric reflection spectra: See Fig.\,\ref{fig.setup}(b) and \ref{fig.setup}(c) and Ref.\,\cite{Schilke:2011} for a detailed explanation of this effect. In practice, the temporal dynamics data presented in the following have been obtained with $\lambda_\lat = 780.80$\,nm, close to the optimum reflection. The trapping depth is then $U_0 \sim 500\,\upmu$K.

After 30\,ms of holding time in the lattice, the atomic sample contains $N \sim 3 - 4 \times 10^8$ atoms at a temperature $T \sim 100\,\upmu$K (measured by time of flight and absorption imaging perpendicular to the lattice), with a density distribution along the lattice that is approximately homogeneous along a length $L \approx 4$\,mm, with Gaussian wings on the edges. Note that this corresponds to $\sim 10^4$ layers. The large number of layers allows us to reach a high reflectivity in spite of the low  refractive index contrast, on the order of a few $10^{-4}$. In the central part of the lattice, the number of atoms per well is $\sim 3\times 10^4$ and the mean density is $\sim 5\times 10^{11}$\,cm$^{-3}$. The ratio $\eta = U_0/\kB T$ is evaluated to $\eta \approx 5$, which gives the size of the Gaussian atomic distribution in each potential well, namely $\sigma_\perp \simeq 100\,\upmu$m and $\sigma_z \simeq 45$\,nm in the transverse and longitudinal directions, respectively. This yields a peak density of $\sim 4\times 10^{12}$\,cm$^{-3}$, well in the dilute regime (average distance between atoms much larger than the wavelength).

The probe beam has a waist $w_0 = 35\,\upmu$m, corresponding to a Rayleigh length of 5.2\,mm and a significant divergence $\Delta\theta=0.4^\circ$. Its intensity is kept weak to avoid any saturation effect. To ensure that all data have been obtained in the linear-optics regime, each measurement pulse is doubled, with the second pulse having five times more power than the first one. We have checked that the results are always the same, apart from the noise level. We show here only the less noisy data, for which the saturation parameter is at most $0.1$. With these parameters, we obtain a maximum Bragg reflection of $R \sim 0.8$ [Fig.\,\ref{fig.setup}(b)], as in \cite{Schilke:2011}.


The light propagation in the system can be modeled by a transfer matrix formalism, as explained in detail in \cite{Deutsch:1995, Artoni:2005, Slama:2006, Schilke:2012b}. In brief, each atomic layer corresponds to a lossy dielectric whose complex refractive index is given by textbook formulas, valid for low density atomic vapors: $n(\omega) = 1+ \rho \alpha(\omega)/2$, where $\rho$ is the atomic density and $\alpha$ the polarizability. Each atomic layer introduces some reflection computed from Fresnel coefficients. To take into account the longitudinal Gaussian distribution of the atoms in each potential well, we actually decompose each layer into several sublayers with a varying density. The total transmitted field through, and reflected field by the whole sample is then obtained by multiplying all transfer matrices. We also take into account the actual distribution of the atomic density along the lattice axis [as seen in Fig.\,S1 of the Supplemental Material (SM) \cite{SuppMat}] and the lattice-induced light shift (as the trapping beams are on during the measurements).

This procedure allows us to compute the output fields (intensity and phase) as a function of the detuning $\delta$ of the incident field. Therefore, we can also compute the temporal dynamics by considering the Fourier transform of the incident pulse, apply the reflection and transmission coefficients in frequency space, and Fourier transform back to temporal space. This is also the method used to compute optical precursors (flash) in the coherent transmission through a disordered medium \cite{Chalony:2011, Kwong:2014, Kwong:2015}. In this procedure, we take into account the finite switch-off duration of the probe beam, which is well described by an erf$(t/\tau)$ profile with $\tau=1$\,ns, and the finite bandwidth of the detector, which is well simulated by applying a Gaussian convolution of rms width 1.3\,ns.

\begin{figure}
\centering\includegraphics{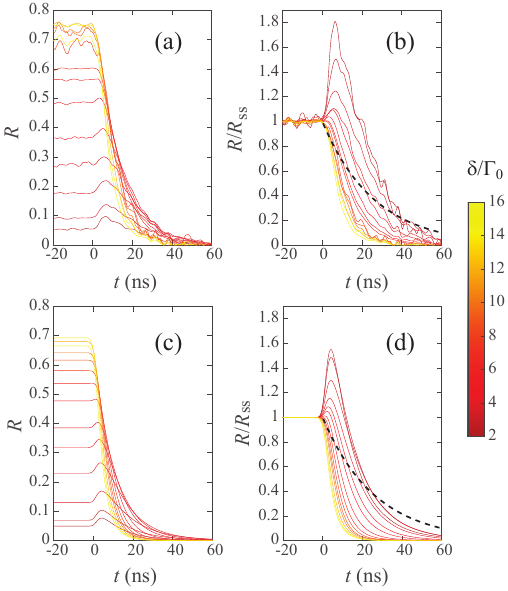}
\caption{(a) Measured reflection coefficient as a function of time at the switch-off of the incident field, for different detunings (indicated in the color bar). (b) Same data but normalized to the steady-state reflectivity $\Rss$, to better see the switch-off dynamics. The black dashed line shows the natural decay, exp$(-t/\tau_\at)$, for comparison. (c, d) Corresponding numerical simulations.}
\label{fig.data}
\end{figure}

\begin{figure}[t]
\centering\includegraphics{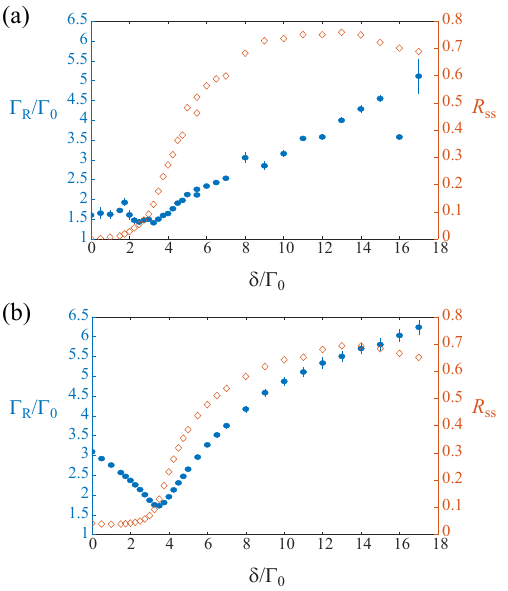}
\caption{(a) Superradiant decay rate of the Bragg reflection as a function of the probe detuning. The error bars are the $\pm 2\sigma$ statistical uncertainties of the fit. On the right axis the steady-state reflection coefficient is reported (measured from the same temporal traces). (b) Simulation with the corresponding parameters. The disagreement for $\delta<3\Gamma_0$ is probably due to the absorption by the remaining atoms from the initial MOT.}
\label{fig.superradiance}
\end{figure}

This simple 1D model needs to be refined for a proper comparison with the experiment, as finite-size effects in the transverse directions are important. Indeed, the finite size of the probe beam induces a significant divergence and the finite size of the atomic cloud changes the atomic density. To include these effects in the 1D model, we average the results over a distribution of rays that samples the angle distribution and the transverse position of the probe beam. For each ray the actual longitudinal density profile is computed and used in the transfer matrix model. The results are quite sensitive to the sizes of the probe beam and the atomic sample. The latter depends on the trapping beam size and temperature. Each number has its uncertainty and, moreover, some imperfections are not included, such as a small astigmatism of the lattice beam, potential unperfect overlap of the counterpropagating lattice beams and imperfect alignment of the probe beam. We have empirically found that reducing the number of atoms by a factor 2 in the model, without any other free parameter, yields good qualitative results and a fairly close quantitative agreement with the experimental data for the steady-state transmission (not shown) and reflection spectra [Figs.\,\ref{fig.setup}(c) and \ref{fig.setup}(d)]. We have kept the same empirical correction for simulating the temporal dynamics without any other free parameter (Figs.\,\ref{fig.data}, \ref{fig.superradiance}, and S2).



Let us now turn to the experimental results on the temporal dynamics. We show in Fig.\,\ref{fig.data} some acquired signals of the reflection as a function of time, zoomed near the switch-off (at $t=0$) of the incident pulse, whose total duration is 0.6\,$\upmu$s (approximately 23\,$\tau_\at$, where $\tau_\at = \Gamma_0^{-1} = 26.2$\,ns), long enough to reach the steady-state. We observe two interesting phenomena. 

First, for some detunings, there is a temporary increase of the reflection. This happens when the detuning is on the resonance side of the reflectivity spectrum (small $\delta$), when the reflectivity is low because of the large amount of scattering losses near resonance \cite{Schilke:2011}. At larger detuning, when the reflectivity becomes higher or when it decreases again on the other side of the spectrum, this flash disappears. The amplitude of the flash is always moderate, on the order of a few percent. However, its \emph{relative} amplitude (compared to the steady-state reflectivity) can be large, when the steady-state reflection is low. In these cases, of course, the data are more noisy. 


We call it ``flash'' by analogy with the phenomenon observed in the coherent transmission through a disordered medium and already studied in detail by Wilkowski \emph{et al.} using a narrow transition of strontium, which is more favorable to resolve very fast transients \cite{Chalony:2011, Kwong:2014, Kwong:2015}. The classical explanation of the flash is the following. The coherent transmission, which can be computed by Beer-Lambert law, is due to the destructive interference between the incident field and the scattered field in the forward transmission. If one is able to switch-off the incident field faster than the time scale at which the scattered field vanishes, one observes a flash, corresponding to the remaining scattered light. In our case, though, this interference picture cannot be used, because the reflected field is only made of scattered light: there is no incident field in the reflection direction. That makes the observed flash in reflection rather counter-intuitive.


In that case, it is easier to find a physical explanation in Fourier space, i.e. considering the frequency broadening induced by switching off the incident field. When the probe detuning is such that the reflectivity is low, but very close to a steep edge in the reflectivity spectrum, the spectral broadening creates frequency components that are better reflected. This yields a transient increase of the reflection, i.e. the flash. The numerical simulations are also based on this principle, and they are in very good agreement with the observations [Figs.\,\ref{fig.data}(c) and \ref{fig.data}(d)]. In SM \cite{SuppMat}, we also discuss an alternative physical picture that happens in temporal space.



Let us now discuss the second, main observation, which is better visible in Fig.\,\ref{fig.data}(b), namely the superradiant decay rate of the reflectivity at the switch-off. To quantitatively study the superradiant dynamics, we fit the observed decay by a single exponential in the range where the intensity relative to the maximum is between 0.8 and 0.2, which works well (all goodness-of-fit parameters $R^2>0.9$, most above 0.99). We also use the same procedure on the simulated data [Fig.\,\ref{fig.data}(d)]. We show the measured decay rates $\Gamma_\mathrm{R}$ as a function of the detuning in Fig.\,\ref{fig.superradiance}, along with the corresponding simulations. We observe a good qualitative agreement between the experimental and numerical data, with a monotonous increase of the decay rate with the detuning (notice that the atomic resonance is light-shifted to $\sim 3.5 \Gamma_0$). We experimentally reach $\GR \sim 5 \Gamma_0$, without reaching any maximum, our measurement being limited by the tunability of the probe laser.

\begin{figure}[t]
\centering\includegraphics{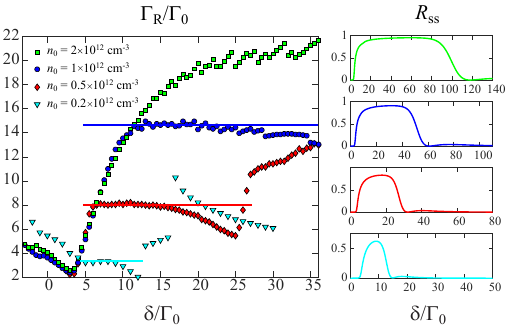}
\caption{Simplified numerical simulations of the decay rate $\GR$ of the Bragg reflection as a function of the detuning $\delta$ for different average densities $n_0$. The corresponding steady-state reflection spectra are plotted on the right. The solid horizontal lines have an horizontal extension that represents the full width at half maximum of the spectra, and their height is adjusted by hand to the data. It gives a ratio $F\!W\!H\!M/\GR = [2.5, 2.8, 3.1]$, for densities $n_0 = [0.2, 0.5, 1]\times 10^{12}$\,cm$^{-3}$, respectively. This approximately constant ratio shows that the decay rate is governed by the width of the spectrum. For the largest density, the decay rate is limited by the extinction rate of the incident laser.}
\label{fig.simu_Gamma_large_delta}
\end{figure}

To investigate what happens at larger detunings, we use our numerical modeling. We ran simplified simulations, i.e. without averaging over the angle or density distribution, as we do not aim at comparing to experimental data, but at understanding the main physics. The results are reported in Fig.\,\ref{fig.simu_Gamma_large_delta} for four different atomic densities, along with the corresponding steady-state spectra. We interpret those results as follows.

When the driving laser is switched off, it creates a spectral broadening. If the different frequency components are not equally reflected, the subsequent narrowing of the spectrum leads to a slower decay compared to the laser switch-off (which is not exponential, but an exponential fit would give $\Gamma_\mathrm{laser} \approx 27 \Gamma_0$). Therefore, the decay rate is maximized when the probe detuning is near the plateau of the reflection spectrum.
In that case, the decay rate is governed by the width of the reflection spectrum. When the detuning approaches the edges of the reflection spectrum, the corresponding strong dispersion yields more distortion and a slower decay rate. This appears very well in the data of Fig.\,\ref{fig.superradiance} and in the simulation of Fig.\,\ref{fig.simu_Gamma_large_delta}. For the largest density, the width of the reflection spectrum is so large that the decay rate starts to be very close to $\Gamma_\laser$. We also observe in Fig.\,\ref{fig.simu_Gamma_large_delta} that, for the two lowest densities, the decay rate starts to increase again when the detuning is beyond the reflection band. This regime of parameters is not very relevant because the reflection coefficient is tiny, but, if it were measurable, the decay rate would indeed tend to $\Gamma_\mathrm{laser}$ at very large detuning, because dispersion effects become negligible far away from any strong spectral feature.

Experimentally, with our moderate densities and detunings, we do not reach the regime of parameters where the decay rate is governed by the width of the reflection. On the contrary, we observe a slight decrease of the decay rate with the atom number. We show this data set in the SM \cite{SuppMat}, along with the corresponding numerical simulations, which are in good agreement. With these parameters, we can conclude that the decay rate is maximal when the spectrum is very broad or flat near the considered detuning. We measured a maximum decay rate of $8 \Gamma_0$.


The observed superradiant decay rate in the Bragg reflection, and the ability of our simple photonic model to explain it, confirms that superradiance in the linear-optics regime is merely a dispersion effect, as already claimed in previous publications dealing with light scattered by disordered samples \cite{Kuraptsev:2017, Guerin:2019, Weiss:2021, Asselie:2022}.
In the case of disordered samples, however, the linear-dispersion model of superradiance discussed in \cite{Weiss:2021, Asselie:2022} neglects interference effects and multiple scattering, and is thus incomplete. It is therefore interesting to check in another system, where interference effects play a major role and incoherent multiple scattering none, that superradiance can still be observed and explained in a similar way.


To summarize, we have studied the extinction dynamics of the Bragg reflection by a 1D lattice of cold atoms. We have observed and explained a flash effect in spite of the absence of incident field in the reflection direction, and a large superradiant decay rate. The two effects can be well reproduced by a 1D photonic model, demonstrating that they essentially are dispersion effects. These findings are an important step towards a comprehensive understanding of such collective effects from the disordered to the ordered case. They also provide a better characterization of this atomic Bragg mirror, useful for future use in photonic applications, such as a two-port all-optical switch \cite{Schilke:2012b}.

Temporal dynamical effects in such a photonic band gap structure is a rich subject and this work opens up different perspectives, which will be the subject of our future investigations. First, it is known that the steep dispersion at photonic band edges can produce very small group velocities (slow light) \cite{Dowling:1994, Scalora:1996} but this has not been observed in atomic systems yet. Second, it will be interesting to measure the decay rate of the fluorescence of atoms driven in the middle of the lattice, and compare the Bragg direction to other directions. Is it superradiant as the Bragg reflection, or is it subradiant, in line with the expected inhibition of spontaneous emission in a photonic band gap \cite{Yablonovitch:1987}?

Finally, let us also note that our hybrid platform features some physical properties that are common to the ones of 1D atomic chains near nanofibers, for example the strong Bragg reflection \cite{Soerensen:2016, Corzo:2016, Olmos:2021}, for which other, more exotic effects have been predicted, for example selected radiance \cite{Asenjo:2017, Pivovarov:2021}. As our sample is easier to produce, it will be interesting to investigate the differences between the two systems. It can also be used more efficiently to study the physics. For example, the enhanced group index at the band edges might be exploited to increase the light-atom coupling in the context of waveguide quantum electrodynamics \cite{Bliokh:2023}.


\emph{Acknowledgements.} We thank Dominique Ronco and Murad Abuzarli for their contributions at the early stage of the experiment and Robin Kaiser for fruitful discussions and constant support. We acknowledge funding from the French National Research Agency through the projects PACE-IN (ANR19-QUAN-003), QuaCor (ANR19-CE47-0014), and 1DOrder (ANR22-CE47-0011). The tunable laser for the lattice was provided by the OPTIMAL Platform.

S.A., J.M.N. and R.C. contributed equally to this work.

\emph{Data availability.} The data are not publicly available. The data are available from the authors upon reasonable request.


%



\clearpage

\onecolumngrid

\LARGE

\begin{center}
\textbf{Supplemental Material:\\
Superradiant decay rate of the Bragg reflection as a function of the atomic density}
\end{center}

\normalsize

\ \\

We show supplementary numerical and experimental data that are not essential to the understanding of the main paper.
\ \\

\noindent\textbf{I. Temporal physical picture of the flash in reflection and intensity in the lattice} \\

In the main paper, we have explained the flash effect in reflection by considering the spectral broadening induced by the switch-off. It is also possible to provide a complementary picture of what happens in time.

At the detunings where the flash appears, i.e. near resonance, the reflectivity is low although the refractive index contrast is high. This is because of the scattering losses, due to the imaginary part of the atomic polarizability, that prevent the wave to penetrate deep into the sample and thus to interact with all the atomic layers. Microscopically, this extinction is due to the interference between the incoming field and the scattered field in the forward direction. When the driving field is switched off, this interference disappears, letting only the scattering field, creating the flash in the forward direction. Moreover, because of the positional order of the atoms in the lattice, a strong constructive interference will also occurs in the backward direction, creating also a flash in the reflection. One could summarize this scenario by saying that the flash in reflection is the reflection of the flash.

\begin{figure}[b]
\includegraphics{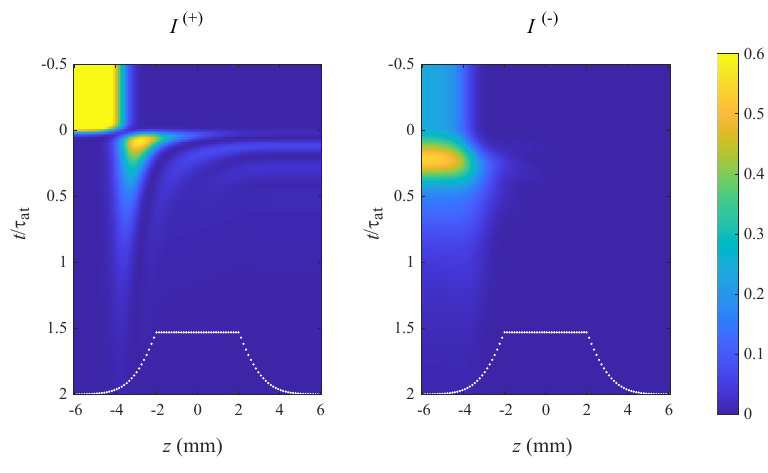}\\
FIG.\,S1. Intensity in the lattice computed with the transfer matrix model. The vertical axis is the time, with the switch-off at $t=0$. The horizontal axis is the longitudinal position within the lattice, whose atomic density distribution is shown as a reference (white dotted line). The left (right) panel shows the wave going in the forward (backward) direction. Here, the detuning is $\delta = 0.5 \Gamma_0$.
\end{figure}

To validate this interpretation, we have run a simplified numerical simulation (i.e., without any averaging over the angle distribution and the transverse density distribution and without the lattice-induced light-shift), based on the transfer matrix formalism, to compute the intensity in the lattice as a function of time. We can compute separately the intensity of the wave going in the forward direction $I^{(+)}$, corresponding to the transmission for $z \rightarrow \infty$, and the wave going in the backward direction $I^{(-)}$, corresponding to the reflection  for $z \rightarrow -\infty$. The result is shown in Fig.\,S1 around the switch-off time $t=0$ ($t$ is the vertical axis). Here, the detuning is close to resonance $\delta = 0.5 \Gamma_0$, such that both the transmission and reflection are low, most of the light being scattered.

In the steady-state ($t<0$), we see that the light hardly penetrates the sample (left panel) and the reflection is also low (right panel). However, just after the switch-off ($t>0$), the wave crosses the sample, creating a flash in the transmission (left panel). This is accompanied by a subsequent increase of the reflection (right panel). This illustrates well our proposed interpretation.


The origin of the oscillating pattern in the transmission is however unclear and we have not observed it experimentally, which might be due to the finite bandwidth of the detector.

\ \\
\ \\

\noindent\textbf{II. Superradiant decay rate of the Bragg reflection as a function of the atomic density}\\

We have studied the superradiant decay rate of the Bragg reflection as a function of the atomic density.
To vary the atomic density, we increase the holding time in the lattice before the probe pulse. This induces a loss of atoms while keeping the lattice length (and atomic distribution) unchanged. The parameter that changes is thus the atomic density. We plot in Fig. S2 as a function of the atom number, which is the parameter that we directly measure.

\begin{figure}[b]
\includegraphics{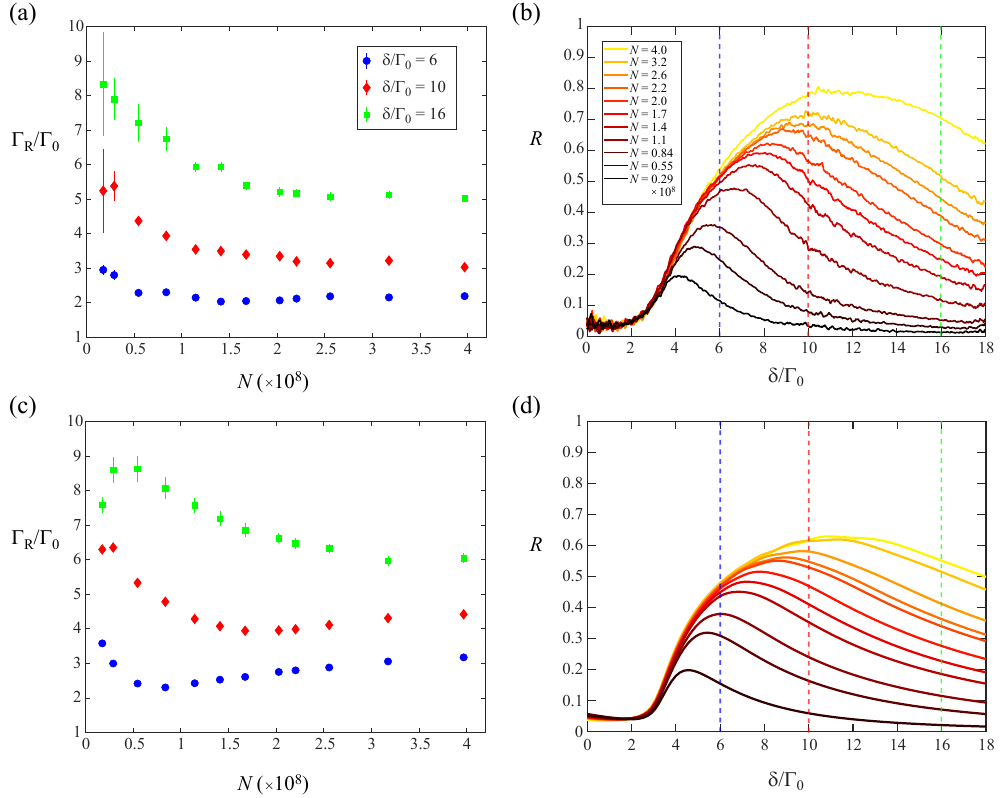}\\
FIG.\,S2. (a) Decay rate $\Gamma_\mathrm{R}$ of the Bragg reflection as a function of the atom number in the lattice, for three different detunings. (b) Corresponding steady-state reflection spectra. (c, d) Corresponding numerical simulations.
\end{figure}

We observe that the decay rate is more superradiant for larger detunings, as reported in the main text [Fig.\,4]. More surprisingly, the general trend is a decrease of the decay rate for an increase of the atom number. The numerical simulations confirm this behavior [Fig.\,S2(c)].

It might seem contradictory with the results reported in Fig.\,5, but in fact, we experimentally do not reach the regime where the decay rate is simply governed by the width of the reflection spectrum. At the moderate detunings and low densities that we have here, the decay rate is given by an intricate interplay between the laser decay rate, the width of the spectrum, and the detuning. Looking at the evolution of the reflection spectrum with increasing atom number [Fig.\,S2(b,d)], we can conclude that the decay rate is larger when the spectrum is broad or flat around the considered detuning, thus inducing less distortion and allowing a faster decay.

Note that superradiant decay rates that decrease with the atom number have already been observed. For instance, for off-axis superradiance in disordered samples in the linear-optics regime, after an initial increase of the decay rate with the resonant optical thickness, the decay rate quickly saturates and then decreases \cite{Araujo:2016}. In a more recent paper \cite{Asselie:2022}, we presented a general framework that interprets linear-optics superradiance as a distortion effect, which mainly depends on the extinction rate of the incident laser, its detuning, and the spectral response of the atomic sample. Therefore, this experiment with ordered samples confirms the generality of the optical picture of superradiance presented in \cite{Asselie:2022} for disordered samples.

\ \\
\ \\

\noindent\textbf{III. Impact of the switch-off duration of the driving field}\\

Following on the previous remark, one can study how the decay dynamics depends on the switch-off duration of the incoming probe beam. In Fig. S3, we present the results of simplified numerical simulations in which the extinction duration is varied. This is parametrized by $\Gamma_\mathrm{laser}$ with the switch-off profile $I(t) = \left[1+\mathrm{erf}(-\Gamma_\mathrm{laser} t)\right]/2$. Experimentaly, we have $\Gamma_\mathrm{laser}^\mathrm{exp} \simeq 26 \Gamma_0$.

As expected, we observe that the flash is maximum for an extremely fast switch-off, and its amplitude decays when the extinction becomes smoother.

Similarly, the superradiant decay rate is also maximum for a very fast extinction, and decreases when $\Gamma_\mathrm{laser}$ is decreased. This is reminiscent of previous results on superradiance in the scattering by disordered samples \cite{Asselie:2022}.

\begin{figure}[h]
\includegraphics{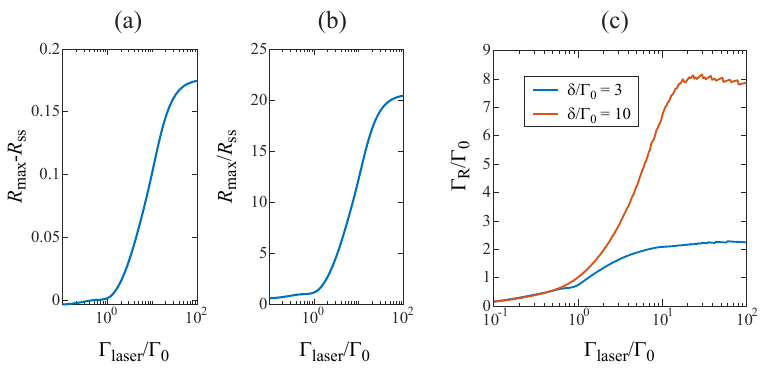}\\
FIG.\,S3. (a, b) Amplitude of the flash as a function of the laser extinction rate $\Gamma_\mathrm{laser}$. In (a), we define the amplitude of the flash as the difference between the maximum reflection coefficient during the switch-off transient response ($R_\mathrm{max}$) and the steady state reflectivity $R_\mathrm{ss}$, and in (b) we plot the \emph{relative} amplitude $R_\mathrm{max}/R_\mathrm{ss}$. (c) Superradiant decay rate of the reflection $\Gamma_\mathrm{R}$ as a function of $\Gamma_\mathrm{laser}$ for two different detunings, one close to resonance ($\delta = 3\Gamma$), where the decay rate is only slightly superradiant, the other at a larger detuning ($\delta = 10\Gamma$), where superradiance is more pronounced.
\end{figure}

\ \\
\ \\

\noindent\textbf{IV. Temporal dynamics at the switch-on}\\

So far, we have focused on the switch-off dynamics. One could also study what happens at the switch-on. With disordered samples, there is also in flash effect in transmission \cite{Chalony:2011} and a collective dynamics in the incoherent scattering \cite{Guerin:2019}.

Unfortunately, due to a technical problem with our pulse generator, we were not able to obtain a clean temporal profile at the switch-on. Therefore, we show in Fig. S4 the results of the same numerical simulations as in Fig. 2(c,d), but now with a closeup on the switch-on dynamics. Since these simulations are in very good agreement with the experimental data at the switch-off (see Fig. 2), we expect them to be also reliable for the switch-on dynamics.

Interestingly, the very beginning of the temporal traces is the same for all detunings, with a fast increase. This is because at very short time, the optical frequency is still not well defined: the spectrum is very broad and therefore the temporal dynamics is fast. Then, when the spectrum narrows as the pulse becomes longer, the behavior becomes detuning-dependent. This behavior is similar to what happens at the switch-off: There is a flash near resonance, and a fast convergence to the steady-state far from resonance.

\begin{figure}[t]
\includegraphics{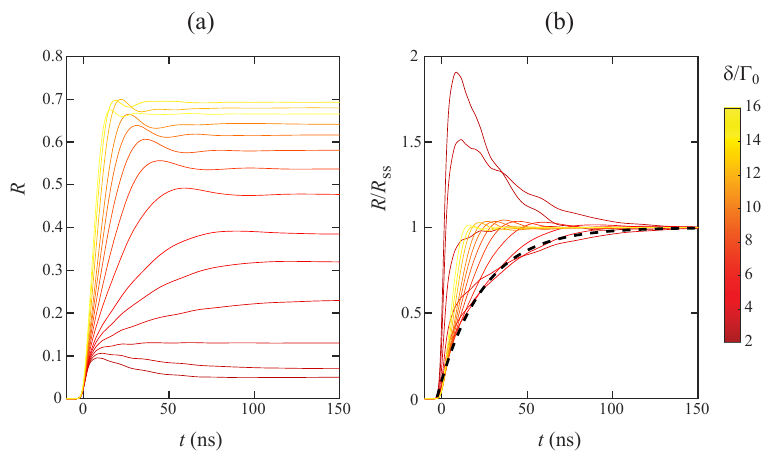}\\
FIG.\,S4. Numerical simulations of the temporal dynamics at the switch-on for different detunings (as indicated in the color bar). (a) Reflection as a function of time. (b) Same data, but the reflection coefficient is normalized to its steady-state value. 
\end{figure}

\end{document}